\documentclass[floatfix,twocolumn,10pt,aps,pre,superscriptaddress]{revtex4}
\usepackage{graphicx}
\usepackage[caption=false]{subfig}
\usepackage{amsmath}
\usepackage{float}

\begin{document}

\title{The determinism and boundedness of self-assembling structures}

\author{S. Tesoro}
\affiliation{Theory of Condensed Matter, Cavendish Laboratory, University of Cambridge, CB3 0HE Cambridge, UK}
\author{S. E. Ahnert}
\affiliation{Theory of Condensed Matter, Cavendish Laboratory, University of Cambridge, CB3 0HE Cambridge, UK}
\affiliation{Sainsbury Laboratory, University of Cambridge, CB2 1LR Cambridge, UK}
\author{A. S. Leonard}
\email{asl47@cam.ac.uk}
\affiliation{Theory of Condensed Matter, Cavendish Laboratory, University of Cambridge, CB3 0HE Cambridge, UK}
\affiliation{Sainsbury Laboratory, University of Cambridge, CB2 1LR Cambridge, UK}

\begin{abstract}
  Self-assembly processes are widespread in nature, and lie at the heart of many biological and physical phenomena.
  The characteristics of self-assembly building blocks determine the structures that they form.
  Two crucial properties are the determinism and boundedness of the self-assembly.
  The former tells us whether the same set of building blocks always generates the same structure, and the latter whether it grows indefinitely. 
  These properties are highly relevant in the context of protein structures, as the difference between deterministic protein self-assembly and nondeterministic protein aggregation is central to a number of diseases.
  Here we introduce a graph theoretical approach that can determine the determinism and boundedness for several geometries and dimensionalities of self-assembly more accurately and quickly than conventional methods.
  We apply this methodology to a previously studied lattice self-assembly model and discuss generalizations to a wide range of other self-assembling systems.
\end{abstract}

\maketitle

\section{Introduction}\label{intro}
Self-assembly is a ubiquitous phenomenon in nature, producing complex structures in biology, chemistry and physics.
Examples include DNA \cite{winfree1998design,goodman2005rapid,rothemund2006folding,ke2012three}, protein quaternary structure \cite{Levy:2006ez,villar2009self}, protein aggregation \cite{Cohen:2013fd}, viruses \cite{lavelle2009phase}, micelles \cite{israelachvili1994self}, and thin films \cite{krausch2002nanostructured}.

Two fundamental questions about a self-assembling system are: do the building blocks always form the same structure and does the assembly grow indefinitely.
The former property is referred to as the {\em determinism} and the latter as {\em boundedness} \cite{Evo,ARXIV2015}.
Together they form the assembly classification.

Protein complexes are a prominent example of deterministic, bound self-assembly.
Misfolding or erroneous binding of mutated versions of such proteins can cause the self-assembly of a protein complex to become nondeterministic, unbound protein aggregation.
This in turn is the hallmark of a number of severe diseases, such as sickle-cell anemia and Alzheimer's \cite{Cohen:2013fd,Bunn:1997de}.

In this paper we introduce a framework for establishing the determinism and boundedness of a given set of self-assembling building blocks.
We apply this approach to a previously studied lattice self-assembly model introduced in \cite{Modularity} and studied further in \cite{Evo,green,greengen,ARXIV2015,TES2016}.
In this model square tiles have attractive interfaces of different types, with interactions governed by a simple set of rules.
The final structures assembled are sets of connected lattice sites known as polyominoes.

Such tile self-assembly or polyomino models have been useful for the study of genotype-phenotype maps, where the specification of the building blocks can be viewed as a genotype and the resulting structure as the phenotype \cite{Evo,green,greengen}.
This approach can be combined with genetic algorithms to model evolutionary processes \cite{Evo}.

Despite being an abstract model, this polyomino model can directly and meaningfully map to real biological self-assembly phenomena.
For example, the sickle-cell mutation of hemoglobin that leads to unbound protein aggregation can be modelled using polyominoes \cite{green}.
Furthermore, experimental implementations of the polyomino model have been realized using DNA tiles \cite{TES2016}. 

The assembly process previously used is fully stochastic, and may be sketched out as: (a) the structure is seeded with a randomly selected tile, (b) a random face on the structure is chosen, (c) a random tile is drawn with a random orientation, and (d) the drawn tile binds to the structure if the interfaces are interacting.
Steps (b-d) are then repeated until no further attachments are possible and assembly terminates.

In this paper, we introduce a graph based approach to replace stochastic assembly as the primary method for identifying the determinism and boundedness of a given set of self-assembly building blocks.
The stochastic approach is inelegant and suffers from a compromise between accuracy and speed, whereas the graph theoretical approach offers a robust methodology that improves accuracy and speed.

This paper proceeds by discussing self-assembly basics in Section \ref{sec:Lsa}. Assembly graphs and their construction from tile sets are introduced in Section \ref{sec:AG}.
Classification preserving transformations of assembly graphs which simplify graph features into deterministic motifs are discussed in Section \ref{sec:ATAP}.
After classifying the assembly graph, the structure is validated under steric constraints in Section \ref{sec:CP}.
Extensions to the method and general discussion are in Sections \ref{sec:FE} and \ref{sec:D} respectively.

With generalizations to other geometries and dimensions, there are potential applications in the study of protein complexes and protein aggregation, as well as bioengineering and nanotechnology.

\section{Lattice self-assembly}\label{sec:Lsa}
Following the model introduced in \cite{Modularity}, a lattice self-assembly tile set consists of one or more tile types.
Each side of a tile (the geometric face) has an interface type, with every tile type in the set exhibiting unique configurations of these interfaces.
Conventionally, interactions are defined with {\bf 0} as non-interacting and {\bf 1} $\leftrightarrow$ {\bf 2}, {\bf 3} $\leftrightarrow$ {\bf 4}, etc. as interacting pairs.
However, these interaction rules can take any arbitrarily complex form provided interactions are bidirectional.
Interactions are infinite in strength, meaning two tiles bind irreversibly if the adjoined interfaces are interacting.
There is an infinite population of each tile type, precluding any stoichiometric limitations.

\subsection{Definitions of structure and determinism}

\begin{figure}[h]
  \centering
  \includegraphics[width=0.45\textwidth]{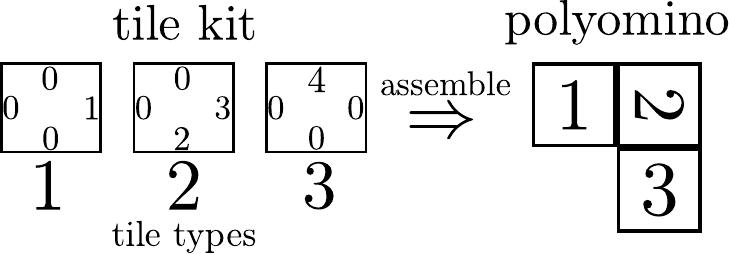}
  \caption{A tile set (left) and its assembled polyomino (right).
    Each tile type in the set can be assigned a number, which is used to label the tile type in the polyomino.
    Rotating the label indicates the corresponding rotation of the tile during assembly.}\label{determinism}
\end{figure}

A structure is defined as a set of connected tiles, each with an associated tile type, orientation, and unique lattice coordinates.
Structure determinism can be defined in three distinct categories, listed in increasing order of strictness: shape, tile, and orientation determinism.
Structures are defined independently of absolute position and rotation, and so lattice coordinate translations or rotations of entire structures are considered indistinguishable (structures are {\em one-sided} polyominoes).
An example of a tile set and its assembled polymino is shown in Figure \ref{determinism}.

{\bf Shape determinism} requires all produced structures to be the same one-sided polyomino.

{\bf Tile determinism} additionally requires corresponding coordinates between two structures to have matching tile types.

{\bf Orientation determinism} further requires matching relative inter-tile orientations.

Orientation determinism is standard choice due to the impact that tile type and relative orientation could have on the function of the assembled structure. 

\section{Assembly graphs}\label{sec:AG}
\begin{figure}[h]
  \centering
  \includegraphics[width=0.475\textwidth]{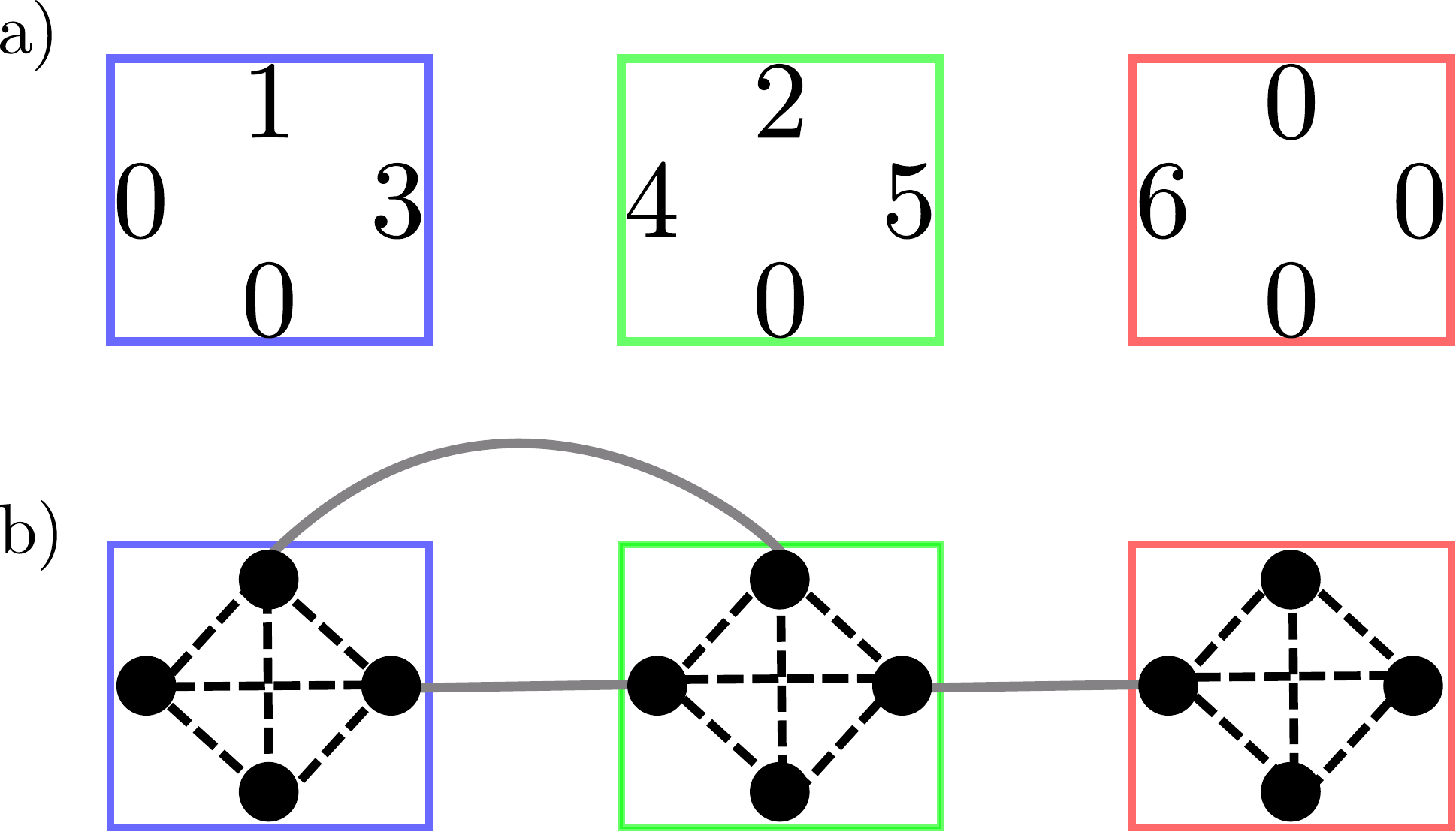}
  \caption{(a) The building blocks for the tile set $\{\left(1,3,0,0\right),\left(2,5,0,4\right),\left(0,0,0,6\right)\}$ explicitly labeled with interfaces and the assembled structure.
  (b) The corresponding complete assembly graph with black (dashed) internal edges, and gray (solid) external edges.}\label{assemblygraphdemo}
\end{figure}

Assembly tile sets are represented using the notation $\{\text{tile type 1}, \text{tile type 2}, \ldots \}$, where tile type ordering is irrelevant.
Tile types have cyclic symmetry and have a length fixed by the geometry, so square tiles have four faces and can be represented as $\left( F_1, F_2,F_3,F_4 \right)$, where $F_i$ indicates the interface type of face $i$.
We use the convention that faces are encoded clockwise starting from the top.
An example of this notation can be found in the caption of Figure \ref{assemblygraphdemo}.

The {\bf assembly graph} of a tile set is constructed with nodes for each face on every tile type.
The faces within each tile type form cliques with edges labeled as internal, while interactions between interfaces (both inter- and intra-tile) are encoded with edges labeled as external. 
Hence an assembly graph can be represented using an edge-labeled pseudograph (multigraph with loops).
A detailed assembly graph example is shown in Figure \ref{assemblygraphdemo}.
Internal edges only depend on geometry, and are not displayed in future examples for clarity.

\subsection{Assembly graph terminology}
\begin{figure}[h]
  \centering
  \includegraphics[width=0.475\textwidth]{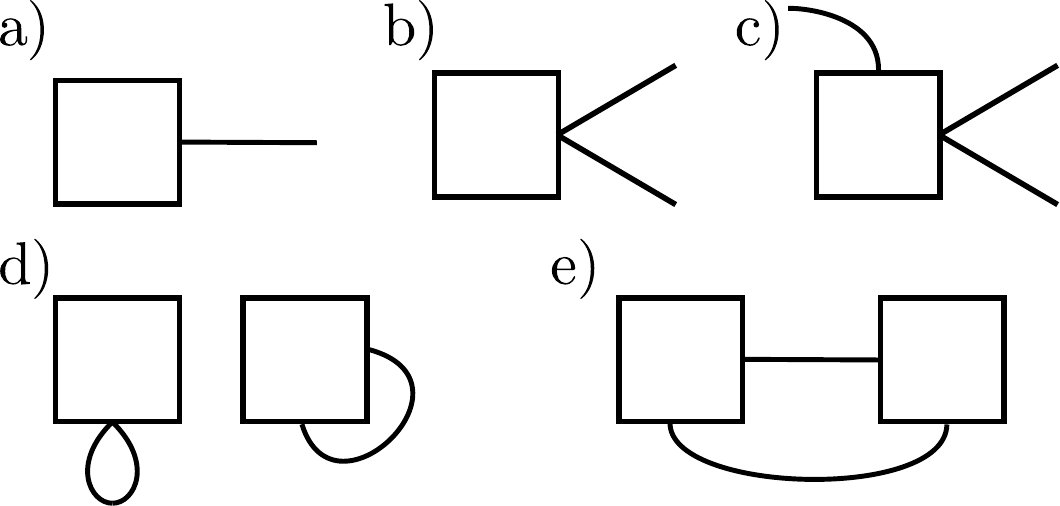}
  \caption{Five partial assembly graphs demonstrating various features.
  (a) SIF tile (b) SIF branching point (c) non-SIF branching point (d) two intra-tile cycles (e) inter-tile cycle.}\label{assemblygraph}
\end{figure}

Assembly graphs contain three features of interest: single interacting faces, branching points, and cycles.
Figure \ref{assemblygraph} shows examples of partial assembly graphs with such features.

{\bf Single interacting face} (SIF) tiles are tile type which have only a single face with one or more external edges.

{\bf Branching points} occur when a given face has multiple external edges.
Non-SIF branching points occur if a tile type has a branching point and at least one other face with external edges.
Non-SIF branching points frequently cause nondeterminism due to diverging assembly pathways.

{\bf Cycles} have two varieties with identical impact on assembly classification: inter- and intra-tile.
Inter-tile cycles are walks on the assembly graph which alternate stepping on external and internal edges.
Intra-tile cycles are walks of only a single step on an external edge connecting to the same tile.
Cycles are the primary source of unboundedness, due to the potential to endlessly traverse (and thus assemble) around the cycle.

\subsection{Graph connectivity}
If an assembly graph contains disconnected components, it is trivially nondeterministic due to seed dependence.
This is due to the inability to assemble between disconnected components (by definition), and hence the structures formed by seeds in different components are nondeterministic.

However, each connected component can be assessed independently for determinism.
Biologically, this can be related to genotypes encoding multiple independent phenotypes.

\subsection{Fundamental Deterministic Graph}\label{subsec:FDG}

A {\bf treelike} assembly graph is one without any cycles or non-SIF branching points.
The simplest treelike graph is a single tile with no interactions, and is evidently bound and deterministic.
Adding a SIF tile (potentially SIF branching point) with a new interaction pair to this assembly graph cannot introduce a non-SIF branching point or cycle by definition.
Since these are the only sources of nondeterminism and unboundedness, the assembly classification consequently cannot be altered.
By induction, any treelike graph is therefore bound and deterministic.

As such, any assembly graph that can be transformed into a treelike graph will also be bound and deterministic.
We now focus on the procedures of reducing an arbitrarily complex assembly graph to this deterministic state, or if this is not possible, identifying the source of nondeterminism or unboundedness.

Sequentially, we prune SIF tiles and check if the assembly graph is treelike.
If we cannot classify the assembly graph at this stage, we analyze the nature of any cycles present and remove trivial cycles, and examine the remaining cycles for unbound behavior.
Although the graph approach identifies sources of nondeterminism and unboundedness, it is unable to predict spatial conflicts that lead to steric nondeterminism, a limitation discussed later.

\section{Alternative treelike assembly procedures}\label{sec:ATAP}

Explicitly transforming assembly graphs to be treelike is unnecessary; the steps of SIF tile pruning to remove trivial branching points and checking for infinite cycles is sufficient to determine the assembly classification.

Assembly graphs which cannot undergo or fail the above procedures are necessarily nondeterministic or unbound.
For instance, any assembly graph with multiple of each complementary interface, i.e.\ branching points connected to branching points, cannot be made treelike and thus is nondeterministic.

\subsection{SIF tiles elimination procedure} \label{subsec:SEP}
SIF tiles, by converse reasoning to that of the fundamental deterministic graph in Section \ref{subsec:FDG}, can be `pruned' from an assembly graph without altering its assembly classification.
This is done by removing a SIF tile from the assembly graph, and neutralizing its complementary interface (or interfaces if it's a SIF branching point).
SIF tiles whose complementary interface is a branching point are nondeterministic and may be classified without pruning.

\begin{figure}[h]
  \centering
  \includegraphics[width=0.425\textwidth]{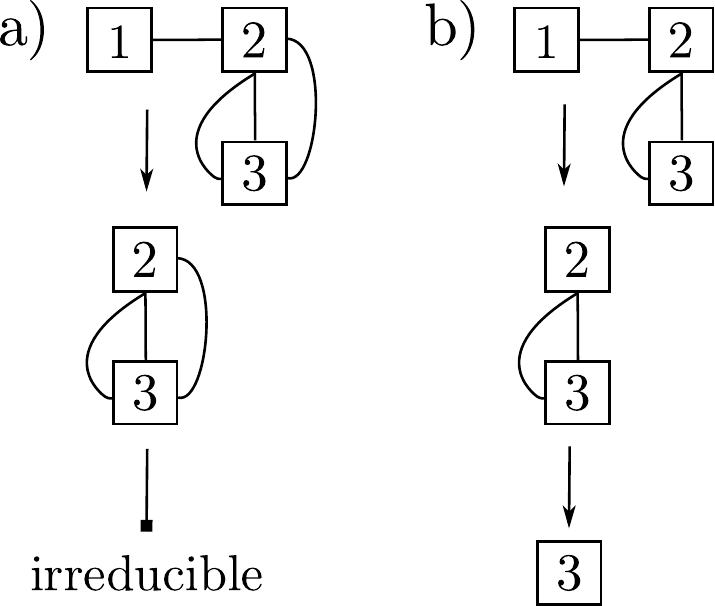}
  \caption{The SIF elimination procedure applied iteratively on two assembly graphs.
    The `1' SIF tile may be pruned in both (a) and (b), but only in (b) does the branching point become a SIF and allow further pruning.
    Note that the second stage of (b) is already treelike (no cycles or non-SIF branching points) and thus classified.
    Since (a) cannot be reduced to treelike, the assembly graph cannot be classified yet.}\label{pruning}
\end{figure}

This procedure is applied iteratively, and if a treelike graph is obtained after a removal, the assembly graph is known to be bound and deterministic.
Examples of the procedure is shown in Figure \ref{pruning}.

\subsection{Cycle rank classification}

\begin{figure}[h]
  \centering
  \includegraphics[width=0.475\textwidth]{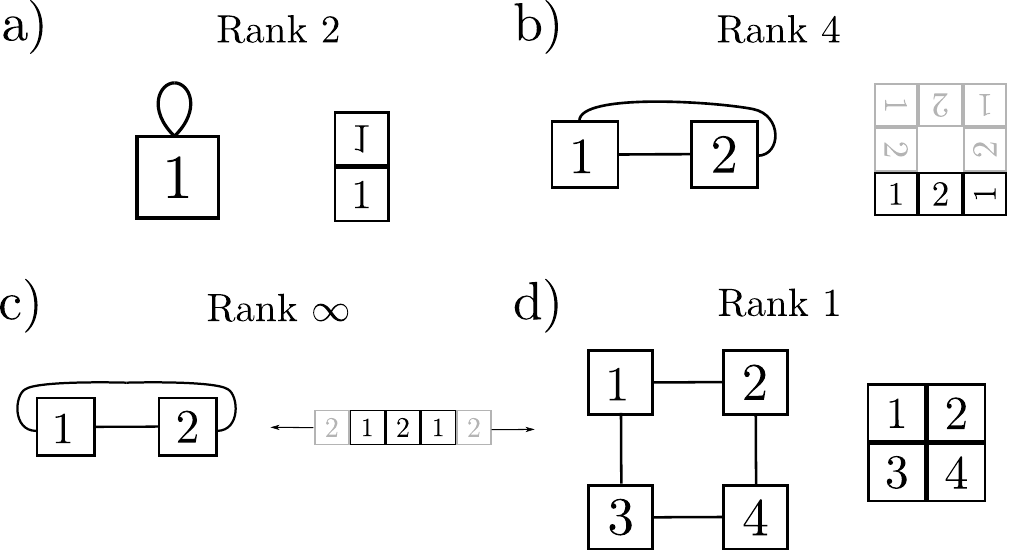}
  \caption{The four cycle ranks possible with square geometry with assembly graphs on the left and assembled structures on the right in each example.
    (a) rank 2 intra-tile cycle ($\Delta \theta=\pi$) (b) rank 4 inter-tile cycle ($\Delta \theta=-\pi/2$) (c) rank $\infty$ inter-tile cycle ($\Delta \theta=\infty, \Delta_{XY} \neq 0$) (d) rank 1 inter-tile cycle ($\Delta \theta=\infty, \Delta_{XY} = 0$).
    As soon as the initially placed tile is reused, the cycle rank may be classified, but for completeness the remaining assembly is shown in gray.
}\label{cycle_rank}
\end{figure}

Cycles can be categorized by the number of times the cycle pattern is repeated during the assembly, a quantity known as the rank.
The rank of the cycle can be determined from a single traversal of the cycle and noting the net rotation accumulated at the end of the walk.
Figure \ref{cycle_rank} shows several examples of cycles and their rank classification.

For square geometry, there are 4 periodic net rotations to consider, $\theta=n \pi / 2$ with $n \in \{-1,0,1,2\}$. 
Rank 4 cycles occur from $\lvert n \rvert=1$, while $n=2$ produces rank 2 cycles.
There are two subcategories for $n=0$, rank $\infty$ and rank 1, based on spatial considerations.

If after traversing the cycle once, there is a zero net spatial translation, i.e.\ the final tile placed reattaches in real-space to the first tile, then the cycle is rank 1.
By contrast, if there is a nonzero net spatial translation and the final tile placed leaves the final interface exposed, then the cycle will grow ad infinitum and is rank $\infty$.

\subsection{Establishing boundedness}

Cycles of rank 1, like SIFs, contribute trivial assembly behavior, and can be simplified without interfering with assembly classification.
Any edge in the cycle may be removed, even if an edge is shared with another cycle.
Regardless of the choice of cut, the resulting assemblies graphs will have the same quantity and rank of surviving cycles.

This procedure is likewise applied iteratively along with SIF elimination until the assembly graph is maximally simplified.
More details are given in Appendix \ref{ap:cycle}.
If at any stage a rank $\infty$ cycle is discovered, growth is immediately known to be unbound.
Moreover, multiple surviving cycles ordinarily exhibit unboundness due to their amalgamated assembly pattern.

\begin{figure}[h]
  \centering
  \includegraphics[width=0.375\textwidth]{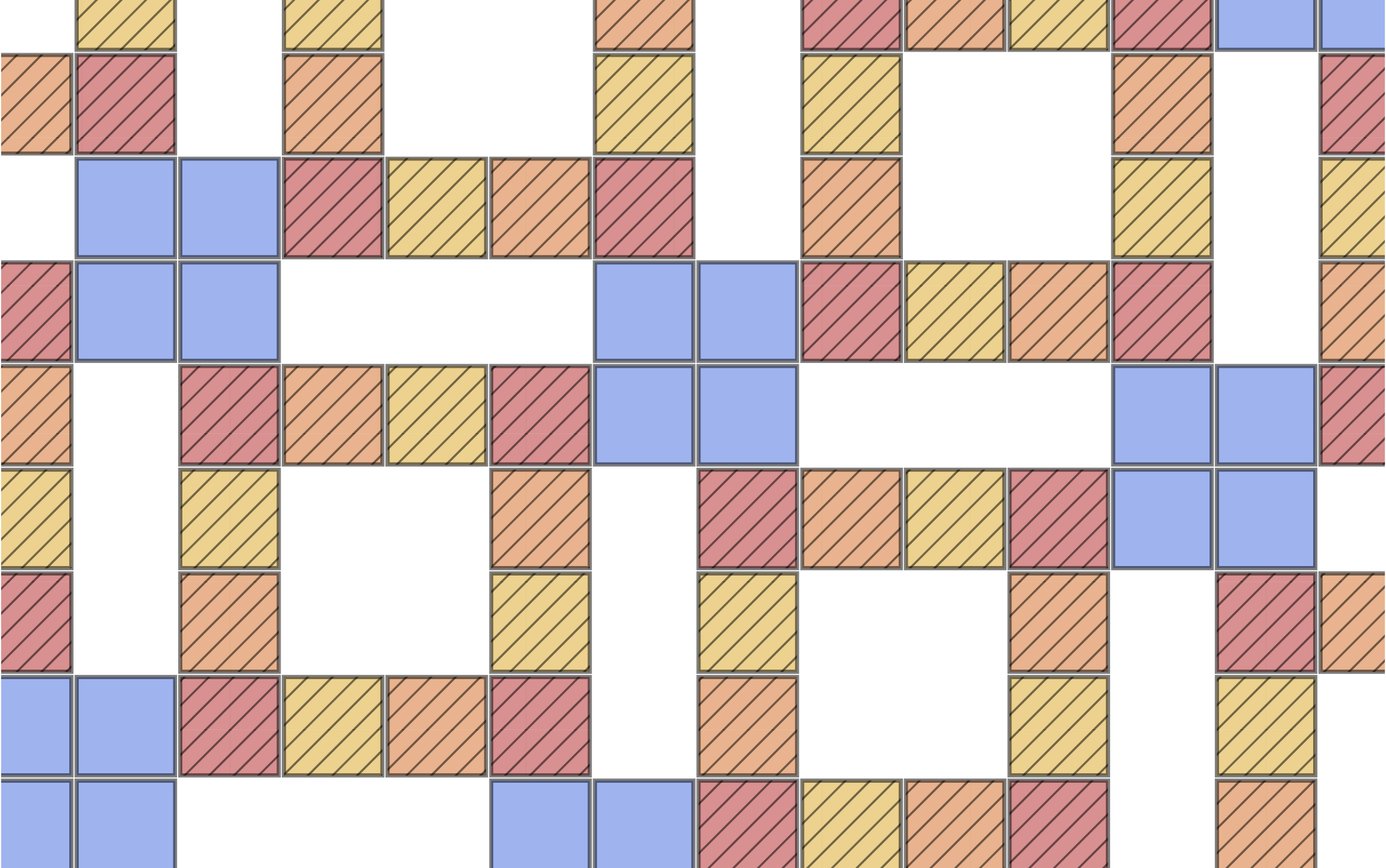}
  \caption{A section of an unbound structure resulting from an assembly graph with two rank 4 cycles.
Every time the blue (plain) cycle is completed, faces which interact with the red-hued (hatched) cycle are left exposed.
A completion of the red-hued cycle likewise allows new copies of the blue cycle to start assembling, leading to unbound growth.}\label{pattern}
\end{figure}

Simplistically, this can be understood as the regeneration of cycle interfaces.
When one cycle completes assembling, interfaces that assemble the other cycles are exposed and available for further growth.
This is demonstrated in Figure \ref{pattern}.

Multiple surviving cycles can have rank 1 behavior rather than rank $\infty$, although it's rare due to the specific spatial constraints required.
This is detailed in Appendix \ref{ap:cycle} for the specific case of Figure \ref{pattern}.

Establishing the (in)finite behavior of the surviving cycles is hence nuanced, and proper care must be taken to identify infinite cycle behavior.
Fundamentally, infinite behavior can be identified if a tile in a cycle of finite rank $R$ is ever used in assembly more than $R$ times.

\section{Complete Procedure}\label{sec:CP}

\subsection{Steric effects}
In the preceding sections we have addressed whether the assembly interactions alone make a tile set (non)deterministic or (un)bound.
However, steric effects can also impact the growth of structures and cause nondeterminism and boundedness in assembly graphs that are otherwise deterministic or unbound.

\begin{figure}[h]
  \centering
  \includegraphics[width=0.35\textwidth]{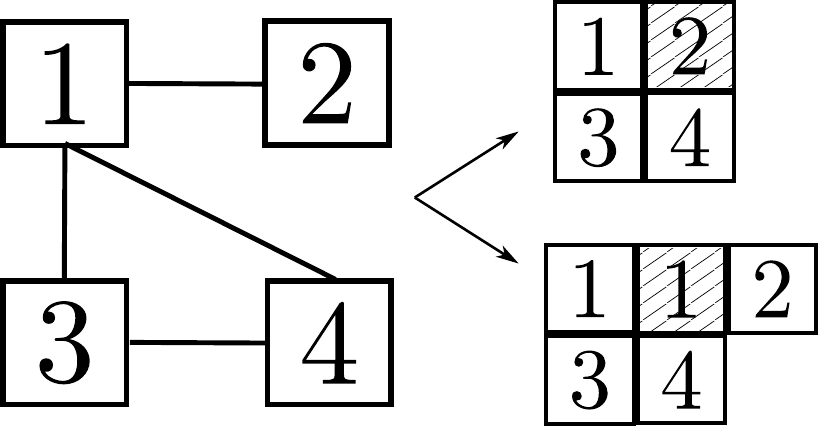}
  \caption{This assembly graph is rule-deterministic, but fails steric validation.
    Growing the structure on the real-space lattice reveals steric nondeterminism as the hatched lattice site does not have a unique occupant.
    This leads to nondeterminism as the final structure would depend on which occupant was built first.}\label{mismatched}
\end{figure}

For this reason a steric validation must be performed by growing the structure on the real-space lattice once.
If any lattice coordinate is used in assembly multiple times, then the assembly is sterically nondeterministic.
Such steric nondeterminism is clearly illustrated in Figure \ref{mismatched}.

Steric nondeterminism can prompt novel behavior in self-limiting cluster growth.
In previous experimental work \cite{TES2016}, single-seed and multi-seed self-assembly were compared, observing that complementary pair interactions can limit growth in multi-seed assembly through local steric effects.
Such phenomenon is beyond the scope of this framework currently, but remains an topic for further analysis.  

\subsection{Geometric symmetry}
The presence of symmetric tiles will never impact classifying a bound and deterministic assembly, but may mistake unbound growth for nondeterminism due to the symmetry-induced branching points.
These spurious nondeterministic branching points can be replaced with alternative interfaces that preserve assembly classification through a desymmetrization procedure (see Appendix \ref{ap:sym} for details).

\subsection{Analysis Sequence}

\begin{figure}[H]
  \centering
  \includegraphics[width=0.45\textwidth]{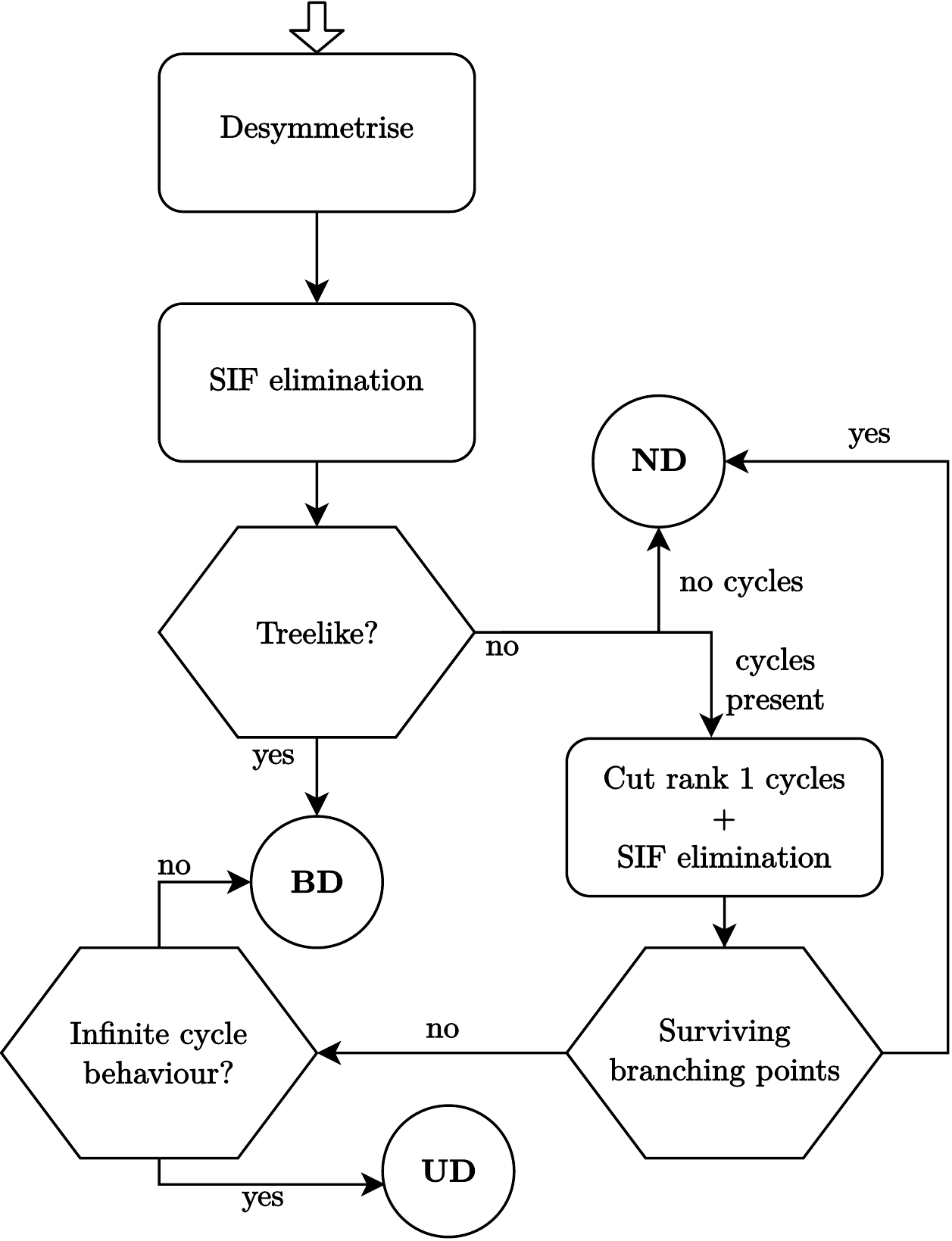}
  \caption{The sequential analysis of an assembly graph.
Classifications can be bound and rule-deterministic (BD), unbound rule-deterministic (UD), or nondeterministic (ND).
Steric validation must still be performed to ensure determinism.
Note disconnected graphs are trivially seed-dependent, but each connected component may undergo this analysis.}\label{flow}
\end{figure}

The flowchart in Figure \ref{flow} illustrates the steps in determining the (un)bound and (non)deterministic nature of a self-assembling tile set. 
An open-source implementation of the assembly graph and steric algorithms is available online\cite{LeonardGithub}.

\section{Framework extensions}\label{sec:FE}
The modular nature of the assembly graph algorithms allows the pinpointing of unbounded and nondeterministic behavior.
Relaxing the conditions which detect the above classifications allows examining a wider range of assembly conditions and constraints.

\subsection{Other dimensions and geometries}
\begin{figure}[h]
  \centering
  \includegraphics[width=0.35\textwidth]{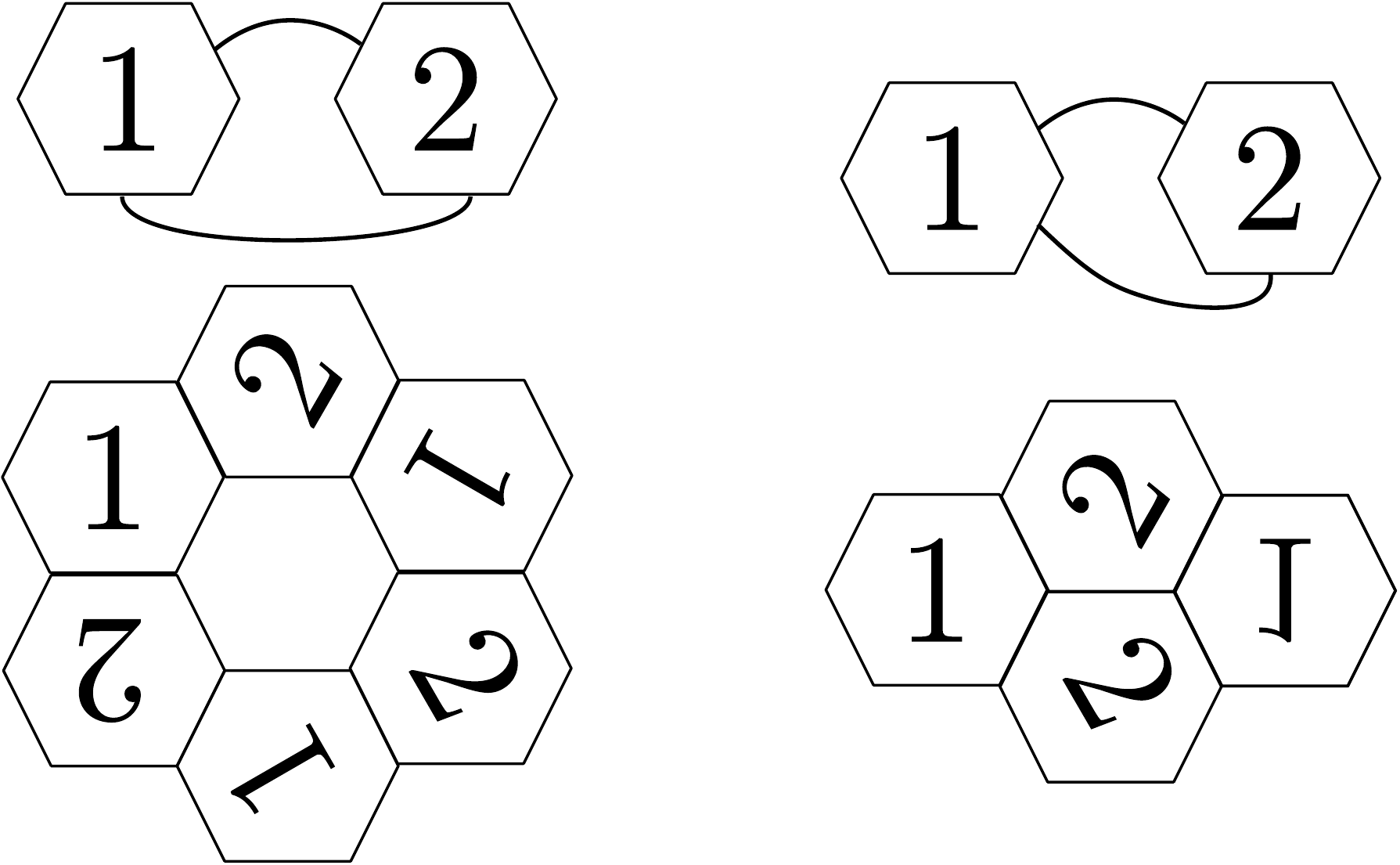}
  \caption{Hexagonal geometry allows for additional cycle ranks.
A rank 3 and rank 2 cycle and the resulting structures are shown on the left and right respectively.}\label{exo}
\end{figure}

The procedures introduced here directly extend to regular triangular and hexagonal geometries, which allow regular tilings of the plane\cite{1977}. 
The only conceptual modification is the $\theta$-shifts in cycle rank classification.
Triangular geometry allows finite cycles of rank 1, 2, and 6, while hexagonal geometry allow ranks of 1, 2, 3, 4, and 6. 
Some cycles examples for regular hexagons are presented in Figure \ref{exo}.

The procedures can also be extended to other dimensions, again only modifying cycle rank classification.
For instance, cubic geometry still only supports finite cycles ranks of 1, 2, and 4, while in one dimension only rank 2 finite cycles are possible.
Although identifying infinite cycle behavior, steric validation, and other mentioned procedures are more complicated to implement, the concepts are unaltered.

\subsection{Nonstatic interaction rules}
Various biological assemblies do not the follow the earlier definition of determinism, due to the ability to form different final structures depending on external conditions, known as {\em phenotype plasticity}.
Such external conditions can considerably impact self-assembly of biomolecules\cite{deamer2006self,lavelle2009phase}.
This framework can encapsulate plasticity by considering e.g.\ high pH interaction rules and low pH interaction rules, where the choice of rules determines which interactions are active.
This in turn can yield different deterministic structures.

Likewise, it is possible to consider sequentially active interaction rules in a single assembly.
An assembly graph constructed under each set of interaction rules can be analyzed for (un)boundness and (non)determinism.
The assembly is then validated under steric constraints by assembling under the first set of interaction rules, followed by using the next set of rules on the existing structure etc.
As such, this approach can generalize to more complex assembly environments.

\subsection{Seed dependence}
Although seed independence is ingrained in the method as outlined, incorporating fixed seed assembly allows for greater diversity in assembled structures.
Seed dependence primarily results from branching points, as the behavior of SIF tiles and cycles are independent of the order of assembly.

\begin{figure}[h]
  \centering
  \includegraphics[width=0.45\textwidth]{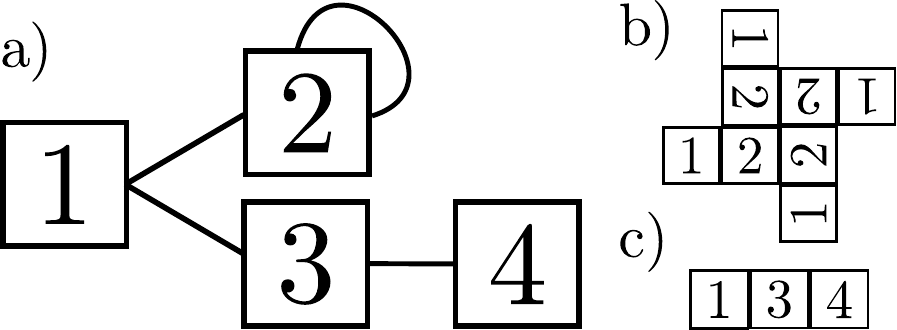}
  \caption{An assembly graph that is classified as nondeterministic.
    However, if the assembly is seeded with tile type `2' or `3'/`4', the structures are deterministic.
    The nondeterministic behavior of the branching point is only encountered walking from the `1' seed.}\label{seeded}
\end{figure}

One possible realization for modifying seed dependence is to walk on the assembly graph depth-first away from the seed.
Branching points where a ``branched to'' face (i.e.\ multiple faces with this interface type exist in the set) is discovered before the complementary ``branching'' face (i.e.\ only face with this interface type in the set) can be ignored.
Intuitively this removes the nondeterminism, as the assembly never encounters the multiple pathways when growing from those seeds.
An illustration of this extension is shown in Figure \ref{seeded}.

\section{Discussion}\label{sec:D}

\subsection{Advantages of the assembly graph framework}
The stochastic assembly method suffers from the probabilistic detection of unboundness and nondeterminism.
Classification accuracy is directly related to the number of repeated assemblies, $K$, which compromises speed.
For example, stochastic assembly can miss the unbounded and nondeterministic nature of the tile set $\{\left(1,0,2,0\right), \left(1,2,0,0\right)\}$ even with seed independence.
The misclassification probability for this tile set has an analytic form for $K$ repeated assemblies given by
\[
\Pr = \sum\limits_{N=1}^{\infty}[N (1/2)^{3+2N}]^K
\]
where $N$ is the number of pairs of the infinite cycle $\left(1,0,2,0\right)$ tiles in the assembly.
For $K=10$, the probability of misclassification is inconsequential ($10^{-15}$).
However, more complex tile sets or weaker determinism definitions can lead to non-negligible loss of accuracy.
Assembling $\{\left(1,2,0,0\right), \left(1,0,0,0\right)\}$ with the less restrictive shape determinism has a misclassification rate of $\left(7/8\right)^{K/2}$, over 50\% for $K=10$.

The form of stochastic assembly algorithm (an example of which is sketched out in Section \ref{intro}) also impacts probabilistic detection.
In the immediately preceding tile set example, randomly drawing a new tile gives equal probability to the two tile types.
However, the $\left(1,2,0,0\right)$ tile type has two potential bindings, and so could also be considered to have twice the probability to be selected over the other tile type.
Depending on the implementation choice, the misclassification rate varies between $\left(19/27\right)^{K/2}$ and $\left(7/8\right)^{K/2}$.
The graph method removes the need for external parameters and model specifications inherent in stochastic assembly.

Due to the exponential growth of configuration space, only one, two, and three tile systems have been exhaustively searched.
However, assembly graphs up to 20 tiles were tested with $10^{11}$ samples per graph size ($N_T$).
All sampled tile sets exhibited no misclassifications when compared with the ``truth'' outcomes of sufficiently reliable ($K=50 N_T$) stochastic assemblies.
A direct comparison of the two methods' speeds is shown in Figure \ref{fig:spm} for connected assembly graphs.
Details on the simulations used can be found in Appendix \ref{ap:speed}.

\begin{figure}[h]
  \centering
  \includegraphics[width=0.475\textwidth]{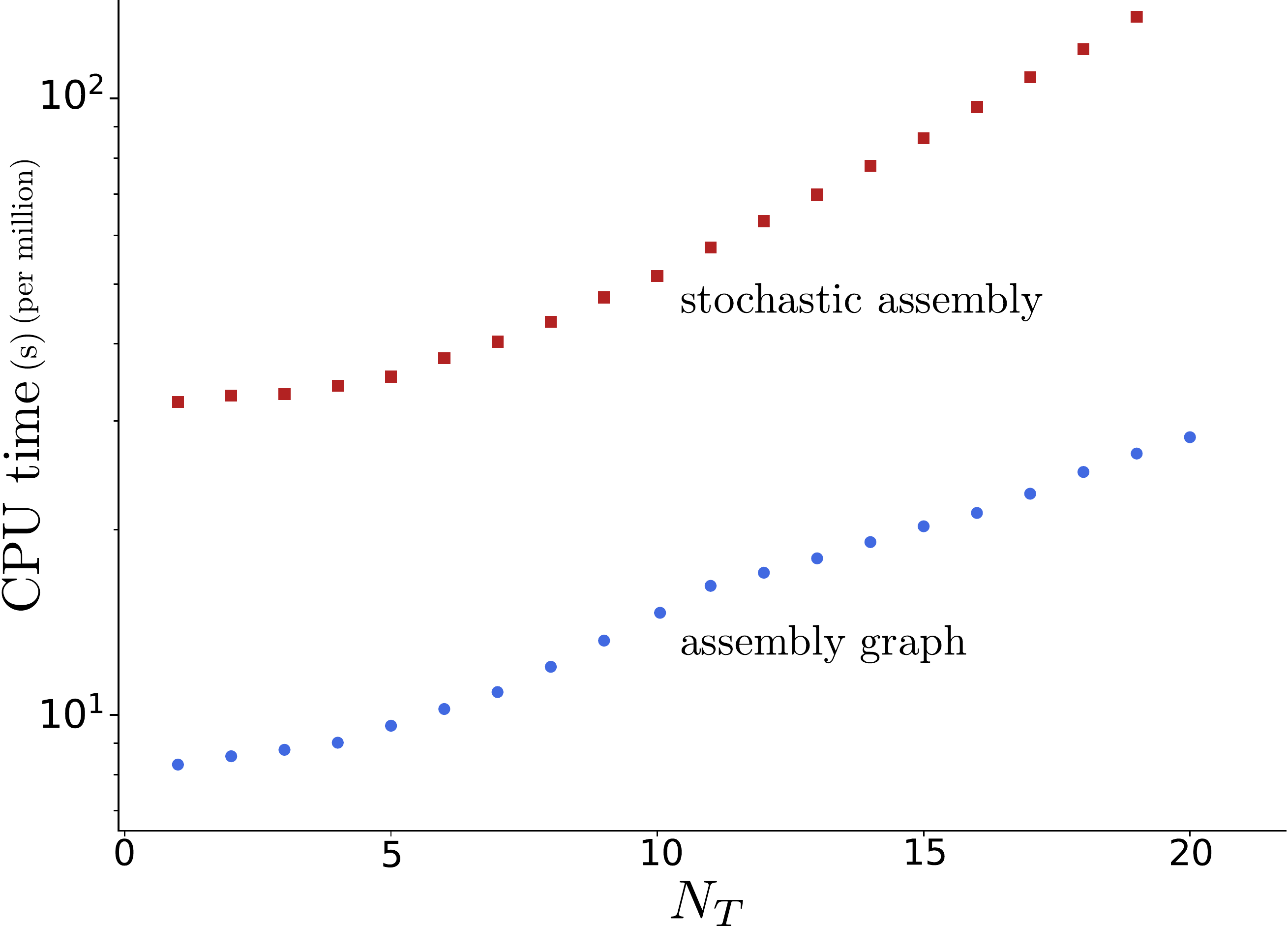}

  \caption{Elapsed CPU time (s) per million connected assembly graphs classified for the graph approach and stochastic assembly with $K=10$.
Both methods initially scale with $N_T$ similarly, although with the graph approach approximately 3 times faster. 
However, for larger $N_T$ the stochastic method scales superlinearly while the graph framework scaling remains approximately linear.}\label{fig:spm}

\end{figure}

\subsection{Nondeterministic assembly dynamics}
In \cite{ARXIV2015}, nondeterministic self-assembly of single lattice tiles and mixtures of two tiles at varying relative concentrations were studied in detail.
In addition to using the conventional interaction rules introduced initially, assembly was also examined using {\bf 1}s and {\bf 2}s as self-interacting.
Tile sets were classified in terms of their behavior upon variation of tile density, recognizing critical transitions from bound to unbound growth and noncritical density transitions.

These dynamics can be analyzed from a different perspective, examining the 106 topologically distinct tile sets with the assembly graph framework.
While identifying deterministic assemblies is relatively unchanged, the ability to robustly distinguish unbound from nondeterministic behavior is greatly improved using the assembly graph framework.

\subsection{Genotype-phenotype maps}

The lattice self-assembly model described here has been used as an abstract model of the genotype-phenotype (GP) map of protein quaternary structure \cite{green}.
In these models, tile set is the genotype and the assembled structure represents the phenotype.
This work focused on bound deterministic assembly, while unbound or nondeterministic building block sets were regarded as a single unfavorable phenotype, and largely ignored.

The assembly graph framework allows assembly graphs to be used as an intermediate link between genotypes and phenotypes, and thus extend such work to examine the GP map of nondeterministic tile sets.

\subsection{Application to real proteins}
The study of such nondeterministic GP maps is essential, as nondeterministic self-assembly in the form of protein aggregation is a hallmark of numerous diseases.
A classic example is that of hemoglobin and the sickle cell mutant.

\begin{figure}[h]
  \centering
  \includegraphics[width=0.4\textwidth]{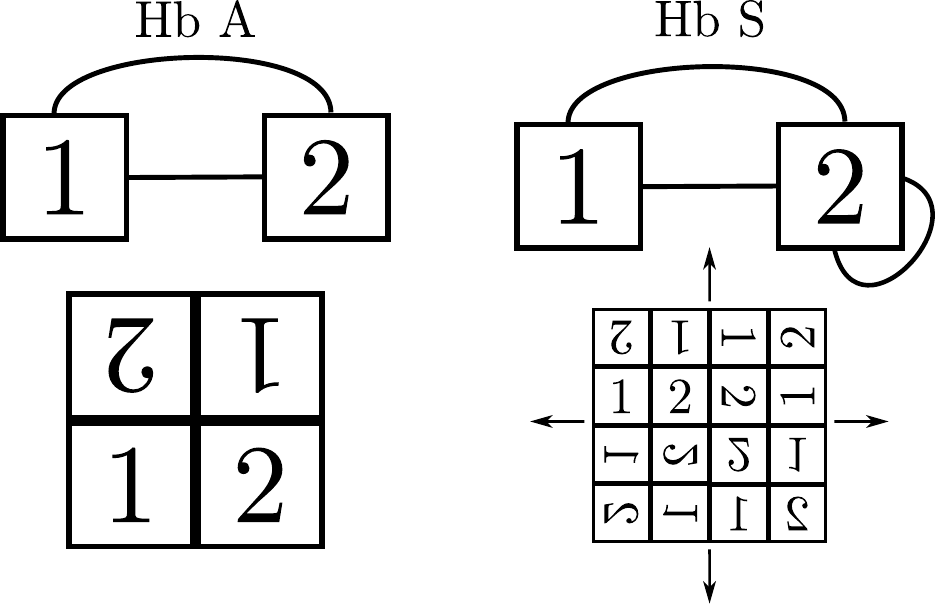}
  \caption{The wild-type (left) and sickle-cell mutant (right) of the hemoglobin complex has been modelled previously using polyominoes\cite{green}.
    The unbound growth denoted by arrows resulting from the mutation can now succinctly be pinpointed to the addition of a higher order cycle to the assembly graph.}\label{GGP}
\end{figure}

In this representation, the $\alpha$ and $\beta$ chains of wild-type hemoglobin Hb A can be mapped to the $\{ \left(1, 3, 0, 0\right), \left(2, 0, 0, 4\right)\}$ tile set, and the sickle-cell mutant Hb S can be mapped to $\{\left(1, 3, 0, 0\right),\left(2, 5, 6, 4\right)\}$, displayed in Figure \ref{GGP}.
The sickle-cell point mutation is sufficient to introduce an interaction labeled here with 5 and 6.
The wild-type assembly graph possesses a cycle of rank 2, and the mutation in the second tile introduces a new cycle of rank 4.
Following the methodology we have outlined, this is sufficient to identify the unbound deterministic growth giving rise to the sickle cell anemia disease.
 
Biological structures that are assembled through cycles with rank greater than 1 are prone to unbound growth via mutations which introduce additional cycles. 
Protein misfolding and resulting changes of interface angles in quaternary structures can be recast as introductions of cycles or branching points.

While many proteins obviously exhibit more complex geometries than this framework is suited to, examining coarse-grained quaternary structure has already revealed insights into assembly properties and evolution history of proteins \cite{Levy:2006ez,ahnert2015principles}.
Combining these protein ``motifs'' with the introduced assembly graph formalisms is a rich vein for future exploration.

\subsection{Universal computing}
The motivations behind this method are unassociated with those of universal computing and Turing machines, but there is conceptual overlap with other tile assembly models (TAMs).
The interaction rules used here define {\em noncooperative} bindings, meaning the interaction is independent of factors beyond the two adjoined faces.
Such noncooperative models have been demonstrated to not intrinsically allow universal computing\cite{meunier2017non}, requiring significant generalizations to the TAMs to increase their computational ability\cite{hendricks2016doubles}.

As such, the framework outlined here for determining boundedness and determinism of self-assembling tile sets has little bearing on universal computation and the infamous halting problem.
  
\section{conclusion}

With the assembly graph framework we have introduced a new approach that can determine the (un)boundedness and (non)determinism of self-assembling tile sets.
While the results presented have focused on 2D square geometry, the assembly framework readily extends to other dimensionalities (1D or 3D) and other geometries which regularly tile the plane. 

The graph based approach outperforms the existing stochastic assembly approach in both speed and accuracy, facilitating the study of significantly larger GP maps and evolutionary dynamics.
External parameter dependencies, like the $K$ repeated assemblies, are also eliminated.

The topological nature of the assembly graphs also opens new options for describing complexity and other properties relevant to evolutionary dynamics.
In addition, this methodology is a powerful tool for generalizing the study of GP maps by extending to the nondeterministic realm, which yields a coarse-grained model relevant to dysfunction and disease in the context of biological self-assembly.

\begin{acknowledgments}
This work was supported by the Engineering and Physical Sciences Research Council (ST and ASL), the Gatsby Foundation (ASL and SEA), and the Royal Society (SEA). 
The authors would like to thank the referees for extensive feedback and suggestions on this work.
\end{acknowledgments}

\appendix

\section{Cycles}\label{ap:cycle}
Determining the rank of a cycle can be done with a single walk around the cycle and tracking the net rotation and translation.
Since each tile has geometry specific cyclic symmetry, the individual steps on the walk contain no information on cycle rank; only the whole walk is useful.
Cycle walks that encounter branching points are almost strictly nondeterministic.
For completeness, these walks can proceed normally, continuing an independent ``cloned'' walk down each branched edge.

There are combinatorial arguments that cycles must have necessarily formed once there are more edges than tiles (accounting for branching points), and the cycles can be detected in many ways.
Exhaustive walks, effectively a depth-first search, on the graph will detect all cycles, or more advanced algorithms can be adapted like Prim's or Kruskal's.

\subsection{Cycle cutting}
\begin{figure}[h]
  \centering
  \includegraphics[width=0.35\textwidth]{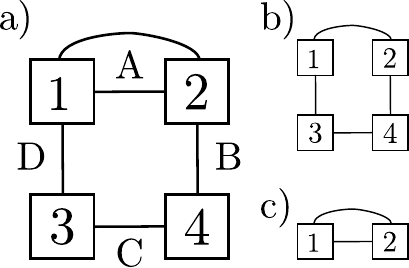}
  \caption{The assembly graph in (a) contains a cycle of rank 1 and one of rank 2.
    Cutting edge A results in the (b) assembly graph, while cutting edge B, C, or D results in the (c) assembly graph.
    Any choice of cut results in an assembly graph with a single rank 2 cycle.
    }\label{cyclecut}
\end{figure}

As soon as a rank 1 cycle is discovered, any single edge in the cycle may be cut without impacting assembly classification.
This is due to the fact that final step in the cycle is unnecessary: the walk returns to the starting point with no net rotation.
An example is shown in Figure \ref{cyclecut}.
While the choice of edge doesn't change the assembly classification, there often is an ``optimal'' choice of cut.
In Figure \ref{cyclecut}, cutting `A' leaves a larger assembly graph, which slows analysis.
However, determining the optimal cut typically involves a similar number of operations as handling the less optimal cut, and so offers little performance gain.

\subsection{Nested cycles}
In general, multiple surviving cycles will grow infinitely.
However, if the multiple cycles are ``nested'', and thus do not regenerate exposed interfaces of the other cycles, then the assembly is bounded.

\begin{figure}[h]
\subfloat{%
  \includegraphics[width=.25\textwidth]{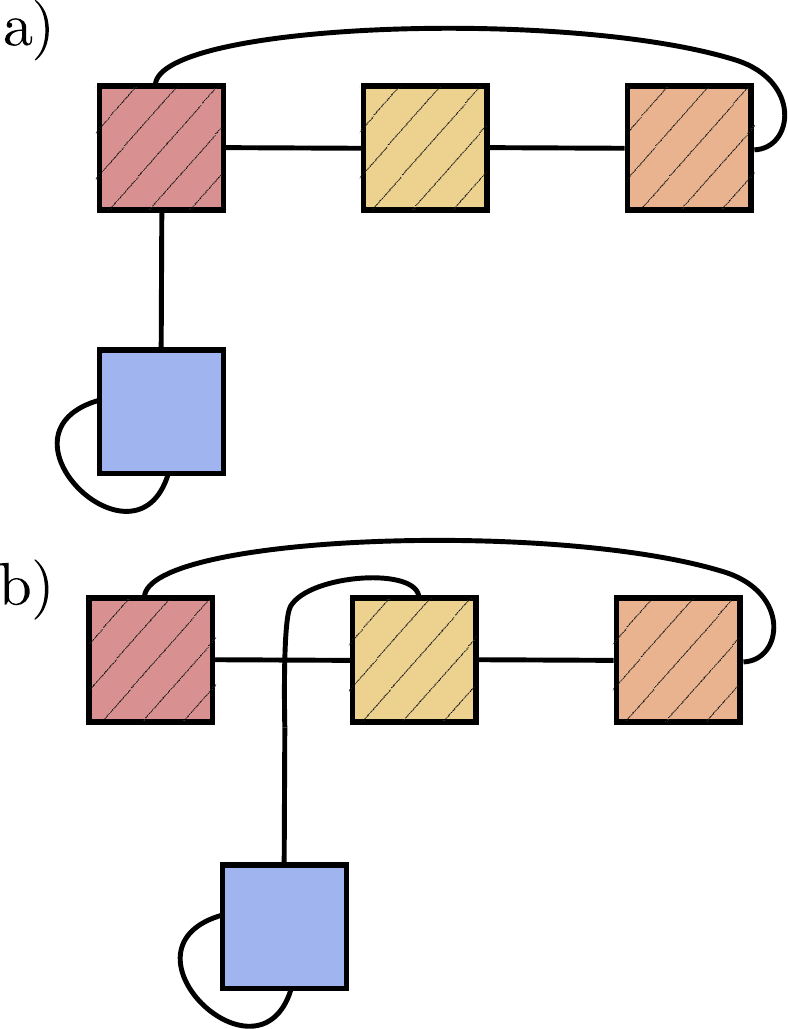}%
}\hfill
\subfloat{%
  \includegraphics[width=.20\textwidth]{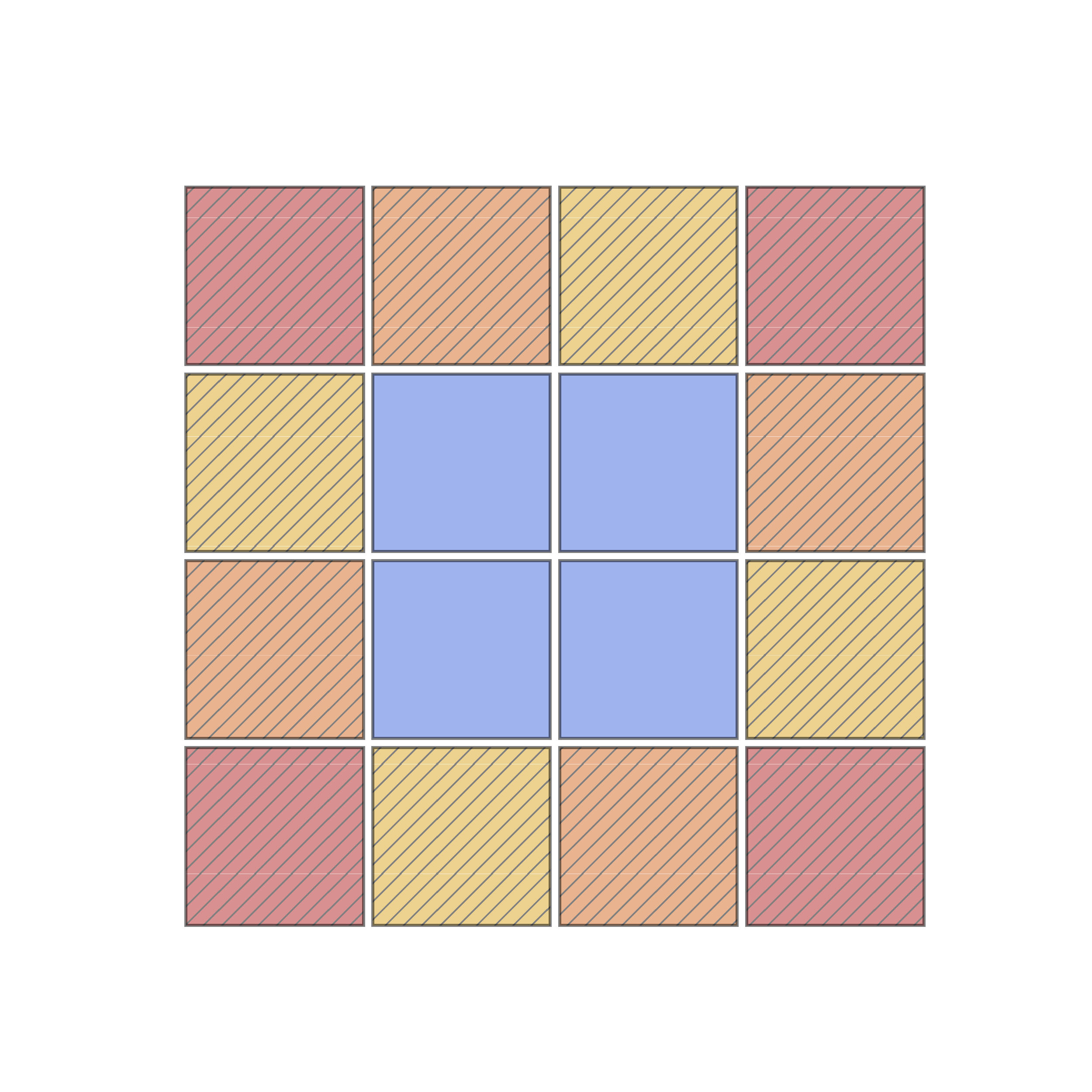}%
}
\caption{The assembly graph for Figure \ref{pattern} is shown in (a), while (b) is a similar assembly graph, but one that is bounded with nested cycle behavior.
  The only difference in assembly graphs is where the tile with the intra-tile cycle joins the other cycle.
  The bounded structure for (b) is shown on the right.
}\label{pattern_ex}
\end{figure}

In the case of Figure \ref{pattern}, this nested behavior can be easily seen by two changes to the tile set.
Relocating where the two cycles connect in the assembly graph, changing $\{\left(1,0,3,2\right),\left({\bf 4},5,7,0\right), \left(0,9,{\bf 0},6\right),\left(0,8,0,10\right)\}$ to $\{\left(1,0,3,2\right),\left({\bf 0},5,7,0\right), \left(0,9,{\bf 4},6\right),\left(0,8,0,10\right)\}$, demonstrates multiple cycles producing rank 1 behavior.
This is shown in Figure \ref{pattern_ex}.
In this case, the completion of one cycle does not generate new exposed faces for the other cycle, and hence precludes infinite growth.

\section{Method comparisons}\label{ap:speed}
The connected assembly graph simulations were comprised of $10^{11}$ samples per graph size ($N_T$), where all sampled assembly graphs had only a single connected component.
This constraint was chosen to prevent e.g.\ a 7 tile graph actually being comprised of 4 and 3 tile disconnected subgraphs.
The number of possible interfaces scaled with tiles as $4 N_{T}+2$ to ensure every possible assembly graph could be sampled, without inflating the neutral space unnecessarily.

The fraction of configuration space which produces unbound or nondeterministic structures increases with the number of tiles.
As such, the accuracy gain drops as the graph size grows since stochastic assembly is able to correctly reject these strongly nondeterministic assembly graphs.
However, the assembly graph approach is still faster and never misclassifies at any $N_T$. 

Evolutionary dynamics from previous work with this model\cite{Evo} were also tested using both methods.
A population of 500 genotypes was evolved under mutation and fitness proportional selection, where the fitness is the size of the assembled structure.
Genotypes were allowed to be disconnected, as each component can be analyzed independently ($K$ assemblies per component), with the maximum fitness across components taken if all were bound and deterministic.

These evolutionary dynamics are most interesting for large configuration spaces (predominantly from large $N_T$).
Since stochastic assembly depends on $K/N_T \gg 1$ for acceptable accuracy, these simulations were either unnecessarily slow for small connected components or inaccurate for large ones.  
The approach introduced here has no fixed external parameters, and so maintains a fast and accurate analysis that can scale with assembly graph size. 

Other simulations were tested, like the nondeterministic assembly dynamics\cite{ARXIV2015}, and had similar behavior to the displayed results with a factor of $5$ speedup and $10^{-6}$ accuracy improvement.

\section{Symmetry removal}\label{ap:sym}
Tiles with geometric symmetry can be substituted with alternative tiles which eliminate symmetry induced branching points but preserve the assembly classification.
For square geometry, symmetric tiles can take one of four forms: $\left(A,A,A,A\right)$, $\left(A,A',A,A'\right)$, $\left(A,B,A,B\right)$, and $\left(A,0,A,0\right)$.
Here $A$ interacts with $A'$ and $B$ has its complementary interface elsewhere in the tile set (not shown).

Many combinations of symmetric tiles are intrinsically nondeterministic, even without considering symmetry induced branching points, and so are not desymmetrized as the analysis will correctly classify these tile sets. 

The type $\left(A,A',A,A'\right)$ can be replaced with $\left(A,A',B,B'\right)$, although this is only useful for a single tile set.
Otherwise, it is nondeterministic if $A$ or $A'$ appear elsewhere in the assembly, and seed-dependent if they don't.

\begin{figure}[h]
  \centering
  \includegraphics[width=0.25\textwidth]{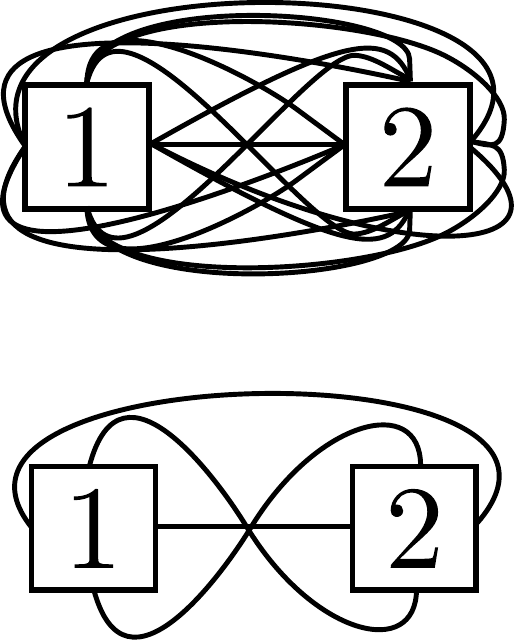}
  \caption{The desymmetrization procedure removes superfluous branching points arising from symmetry that don't contribute unique assembly behaviour.
    The top assembly graph is desymmetrized to yield the bottom one, which preserves the unbound deterministic assembly.}\label{SSS}
\end{figure}

The combination $\{\left(A,B,A,B\right), \left(A',B',A',B'\right) \}$ assembles identically to $\{ \left(A,A,A,A\right),\left(A',A',A',A'\right) \}$ which was already treated in the main text as $\{ \left(A,B,C,D\right),\left(A',B',C',D'\right)\}$.
This is clearly seen a the tile set e.g\ $\{ \left(1,1,1,1\right),\left(2,2,2,2\right)\}$, which is transformed to tile set $\{ \left(1,3,5,7\right),\left(2,4,6,8\right) \}$ in Figure \ref{SSS}.
This case can be ignored if there are more than two tiles in the set as the classification is nondeterministic or seed-dependent if any of the interfaces do or do not appear elsewhere in the tile set.

The remaining cases we must consider are the $\left(A,A,A,A\right)$ and $\left(A,B,A,B\right)$ forms combined with $\left(A',0,0,0\right)$, $\left(A',0,A',0\right)$, and $\left(A',A',A',A'\right)$.
Combinations containing $\left(A',A',0,0\right)$ and $\left(A',A',A',0\right)$ are always nondeterministic and are ignored.

The cases $\{ \left(A,A,A,A\right),\left(A',0,0,0\right)\}$ and $\{ \left(A,B,A,B\right),\left(A',0,0,0\right)\}$ can be ignored as well, because they are either able to be reduced via the SIF elimination procedure or nondeterministic otherwise.
The case $\{\left(A,A,A,A\right),\left(A',0,A',0\right)\}$ requires the addition of a third tile to preserve classification, and is treated as $\{\left(A,B,C,D\right),\left(0,D',0,B'\right),\left(0,C',0,A'\right)\}$.
Finally the case $\{\left(A,0,A,0\right),\left(A',0,A',0\right)\}$ can be replaced with $\{\left(A,0,B,0\right),\left(A',0,B',0\right)\}$.

\begin{figure}[h]
  \centering
  \includegraphics[width=.475\textwidth]{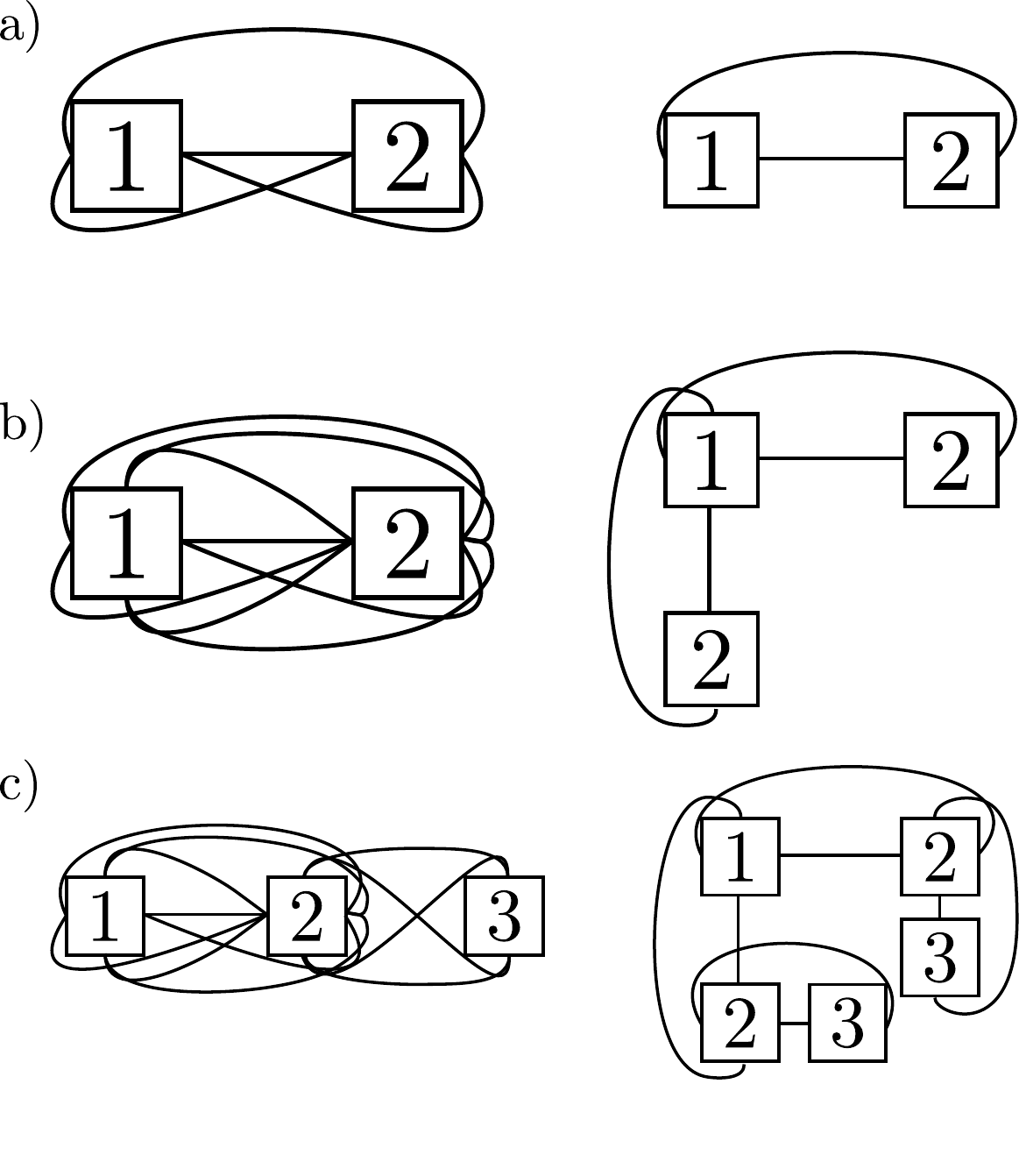}
  \caption{Three example assembly graphs (left) which can be desymmetrized (right) as described above.
    (a) and (b) are single desymmetrizations described in the text, while (c) combines both desymmetrization rules used in (a) and (b) into a single assembly graph.
    Introduced tiles are labeled with the original tile label to ensure the desymmetrized assembly is indiscernible from the original assembly. }\label{ap:symrem}
\end{figure}

In the event of multiple symmetric tiles, the above rules may be combined sequentially to eliminate all symmetry.
If an existing symmetry is unable to be eliminated, then the assembly is nondeterministic already as it will strictly not have a treelike expansion.
Some of these rules and an example of combined desymmetrization are shown in Figure \ref{ap:symrem}.

In all cases, introduced tiles should be considered ``indistinguishable'' from the tile it desymmetrizes.
This is to prevent any tile or orientation determinism issues during the steric assembly, as they are not actually different tiles but rather a bookkeeping tool.
This is shown in Figure \ref{ap:symrem} by labeling introduced tiles with the original tile label.

\bibliographystyle{unsrt}

\bibliography{PRreferences}                                 
\end{document}